\documentclass[twoside,twocolumn,9pt]{article}
\usepackage{extsizes}
\usepackage[super,sort&compress,comma]{natbib} 
\usepackage[version=3]{mhchem}
\usepackage[left=1.5cm, right=1.5cm, top=1.785cm, bottom=2.0cm]{geometry}
\usepackage{balance}
\usepackage{times,mathptmx,wasysym}
\usepackage{sectsty}
\usepackage{graphicx} 
\usepackage{lastpage}
\usepackage[format=plain,justification=justified,singlelinecheck=false,font={stretch=1.125,small,sf},labelfont=bf,labelsep=space]{caption}
\usepackage{float}
\usepackage{fancyhdr}
\usepackage{fnpos}
\usepackage[english]{babel}
\addto{\captionsenglish}{%
  
}
\usepackage{array}
\usepackage{droidsans}
\usepackage{charter}
\usepackage[T1]{fontenc}
\usepackage[usenames,dvipsnames]{xcolor}
\usepackage{setspace}
\usepackage[compact]{titlesec}
\usepackage{hyperref}

\usepackage{siunitx}
\usepackage{subcaption}
\usepackage{amsmath}
\usepackage{amssymb}

\usepackage{epstopdf}

\definecolor{cream}{RGB}{222,217,201}

\begin{document}

\pagestyle{fancy}
\thispagestyle{plain}
\fancypagestyle{plain}{

\fancyhead[C]{\includegraphics[width=18.5cm]{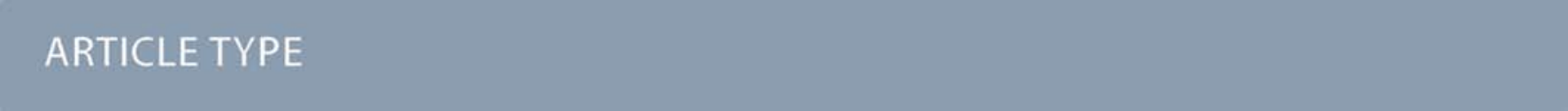}}
\fancyhead[L]{\hspace{0cm}\vspace{1.5cm}\includegraphics[height=30pt]{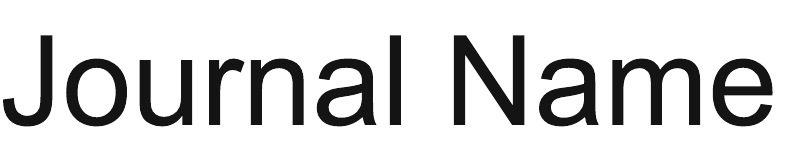}}
\fancyhead[R]{\hspace{0cm}\vspace{1.7cm}\includegraphics[height=55pt]{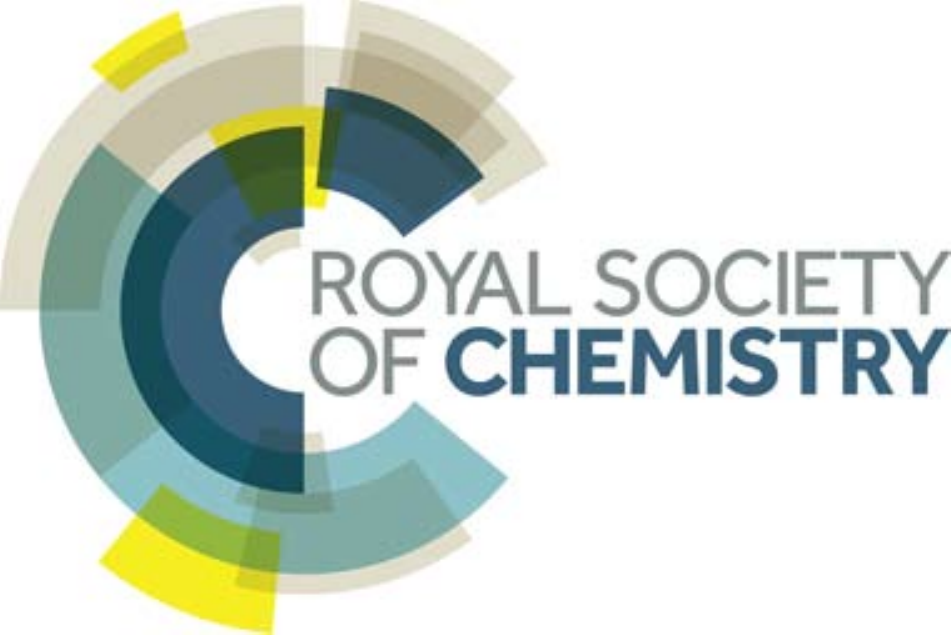}}
\renewcommand{\headrulewidth}{0pt}
}

\makeFNbottom
\makeatletter
\renewcommand\LARGE{\@setfontsize\LARGE{15pt}{17}}
\renewcommand\Large{\@setfontsize\Large{12pt}{14}}
\renewcommand\large{\@setfontsize\large{10pt}{12}}
\renewcommand\footnotesize{\@setfontsize\footnotesize{7pt}{10}}
\makeatother

\renewcommand{\thefootnote}{\fnsymbol{footnote}}
\renewcommand\footnoterule{\vspace*{1pt}%
\color{cream}\hrule width 3.5in height 0.4pt \color{black}\vspace*{5pt}} 
\setcounter{secnumdepth}{5}

\newcommand{\red}[1]{{\color{red}#1}}
\newcommand{\blue}[1]{{\color{blue}#1}}
\newcommand{\gdot}[1]{\dot{\gamma}#1}

\makeatletter 
\renewcommand\@biblabel[1]{#1}            
\renewcommand\@makefntext[1]%
{\noindent\makebox[0pt][r]{\@thefnmark\,}#1}
\makeatother 
\renewcommand{\figurename}{\small{Fig.}~}
\sectionfont{\sffamily\Large}
\subsectionfont{\normalsize}
\subsubsectionfont{\bf}
\setstretch{1.125} 
\setlength{\skip\footins}{0.8cm}
\setlength{\footnotesep}{0.25cm}
\setlength{\jot}{10pt}
\titlespacing*{\section}{0pt}{4pt}{4pt}
\titlespacing*{\subsection}{0pt}{15pt}{1pt}

\fancyfoot{}
\fancyfoot[LO,RE]{\vspace{-7.1pt}\includegraphics[height=9pt]{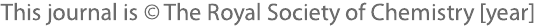}}
\fancyfoot[CO]{\vspace{-7.1pt}\hspace{13.2cm}\includegraphics{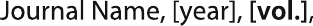}}
\fancyfoot[CE]{\vspace{-7.2pt}\hspace{-14.2cm}\includegraphics{head_foot/RF}}
\fancyfoot[RO]{\footnotesize{\sffamily{1--\pageref{LastPage} ~\textbar  \hspace{2pt}\thepage}}}
\fancyfoot[LE]{\footnotesize{\sffamily{\thepage~\textbar\hspace{3.45cm} 1--\pageref{LastPage}}}}
\fancyhead{}
\renewcommand{\headrulewidth}{0pt} 
\renewcommand{\footrulewidth}{0pt}
\setlength{\arrayrulewidth}{1pt}
\setlength{\columnsep}{6.5mm}
\setlength\bibsep{1pt}

\makeatletter 
\newlength{\figrulesep} 
\setlength{\figrulesep}{0.5\textfloatsep} 

\newcommand{\topfigrule}{\vspace*{-1pt}%
\noindent{\color{cream}\rule[-\figrulesep]{\columnwidth}{1.5pt}} }

\newcommand{\botfigrule}{\vspace*{-2pt}%
\noindent{\color{cream}\rule[\figrulesep]{\columnwidth}{1.5pt}} }

\newcommand{\dblfigrule}{\vspace*{-1pt}%
\noindent{\color{cream}\rule[-\figrulesep]{\textwidth}{1.5pt}} }

\makeatother

\twocolumn[
  \begin{@twocolumnfalse}
\vspace{3cm}
\sffamily
\begin{tabular}{m{4.5cm} p{13.5cm} }

\includegraphics{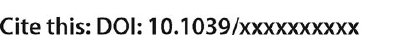} & \noindent\LARGE{Testing the Wyart-Cates model for non-Brownian shear thickening using bidisperse suspensions$^\dag$} \\
\vspace{0.3cm} & \vspace{0.3cm} \\

 & \noindent\large{Ben M. Guy,\textit{$^{a}$} Christopher Ness,\textit{$^{b,c}$}, Michiel Hermes,\textit{$^{a,d}$} Laura J. Sawiak,\textit{$^{a}$} Jin Sun,\textit{$^{b}$}  and Wilson C. K. Poon\textit{$^{a}$}} \\

\includegraphics{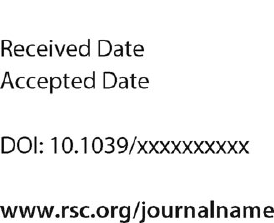} & \noindent\normalsize{There is a growing consensus that shear thickening of concentrated dispersions is driven by the formation of stress-induced frictional contacts. The Wyart-Cates (WC) model of this phenomenon, in which the microphysics of the contacts enters solely via the fraction $f$ of contacts that are frictional, can successfully fit flow curves for suspensions of weakly polydisperse spheres. However, its validity for ``real-life'', polydisperse suspensions has yet to be seriously tested. By performing systematic simulations on bidisperse mixtures of spheres, we show that the WC model applies only in the monodisperse limit and fails when substantial bidispersity is introduced. We trace the failure of the model to its inability to distinguish large-large, large-small and small-small frictional contacts. By fitting our data using a polydisperse analogue of $f$ that depends separately on the fraction of each of these contact types, we show that the WC picture of shear thickening is incomplete. Systematic experiments on model shear-thickening suspensions corroborate our findings, but highlight important challenges in rigorously testing the WC model with real systems.
Our results prompt new questions about the microphysics of thickening for both monodisperse and polydisperse systems.
} \\

\end{tabular}

 \end{@twocolumnfalse} \vspace{0.6cm}

  ]
\renewcommand*\rmdefault{bch}\normalfont\upshape
\rmfamily
\section*{}
\vspace{-1cm}


\footnotetext{\textit{$^{a}$~School of Physics and Astronomy, University of Edinburgh, King's Buildings, Peter Guthrie Tait Road, Edinburgh EH9 3FD.}}
\footnotetext{\textit{$^{b}$~School of Engineering, University of Edinburgh, King's Buildings, Peter Guthrie Tait Road, Edinburgh EH9 3JL}}
\footnotetext{\textit{$^{c}$~Department of Chemical Engineering and Biotechnology, University of Cambridge, Cambridge CB3 0AS, UK.}}
\footnotetext{\textit{$^{d}$~Soft Condensed Matter, Debye Institute for Nanomaterials Science, Utrecht University, Princetonplein 5, 3584 CC Utrecht, The Netherlands}}

\setcounter{footnote}{1}
\footnote{Electronic Supplementary Information (ESI) available: for details on experimental methods and artefacts; an example of complex shear thickening of monodisperse spheres; and fits of the extended WC model for all our simulated binary mixtures. See DOI: 10.1039/b000000x/}




\section{Introduction}
Shear thickening, the increase in viscosity $\eta$ with shear stress $\sigma$ or rate $\gdot$, is ubiquitous in concentrated suspensions\cite{barnes1989shear}. Its microscopic origin has been hotly debated\cite{brown2014shear}. Recent experiments\cite{guy2015towards,lin2015hydrodynamic,royer2016rheological,comtet2017pairwise,clavaud2017revealing}, simulations\cite{seto2013discontinuous,mari2014shear} and theoretical modelling\cite{wyart2014discontinuous} point to a $\sigma$-dependent transition from frictionless (sliding) to frictional (rolling) inter-particle contacts. A phenomenological model by Wyart and Cates\cite{wyart2014discontinuous} (WC) predicts thickening based on a single microphysical parameter, the fraction of frictional contacts, $f$. It fits well the rheology of model systems\cite{guy2015towards,royer2016rheological}; however, its validity for complex industrial suspensions remains untested. We systematically explore the conditions under which the WC model breaks down for one kind of complexity: size polydispersity, and reveal important shortcomings in our current understanding of shear thickening.

\setcounter{footnote}{2}

\begin{figure}[t]
\centering
\includegraphics[width=\columnwidth]{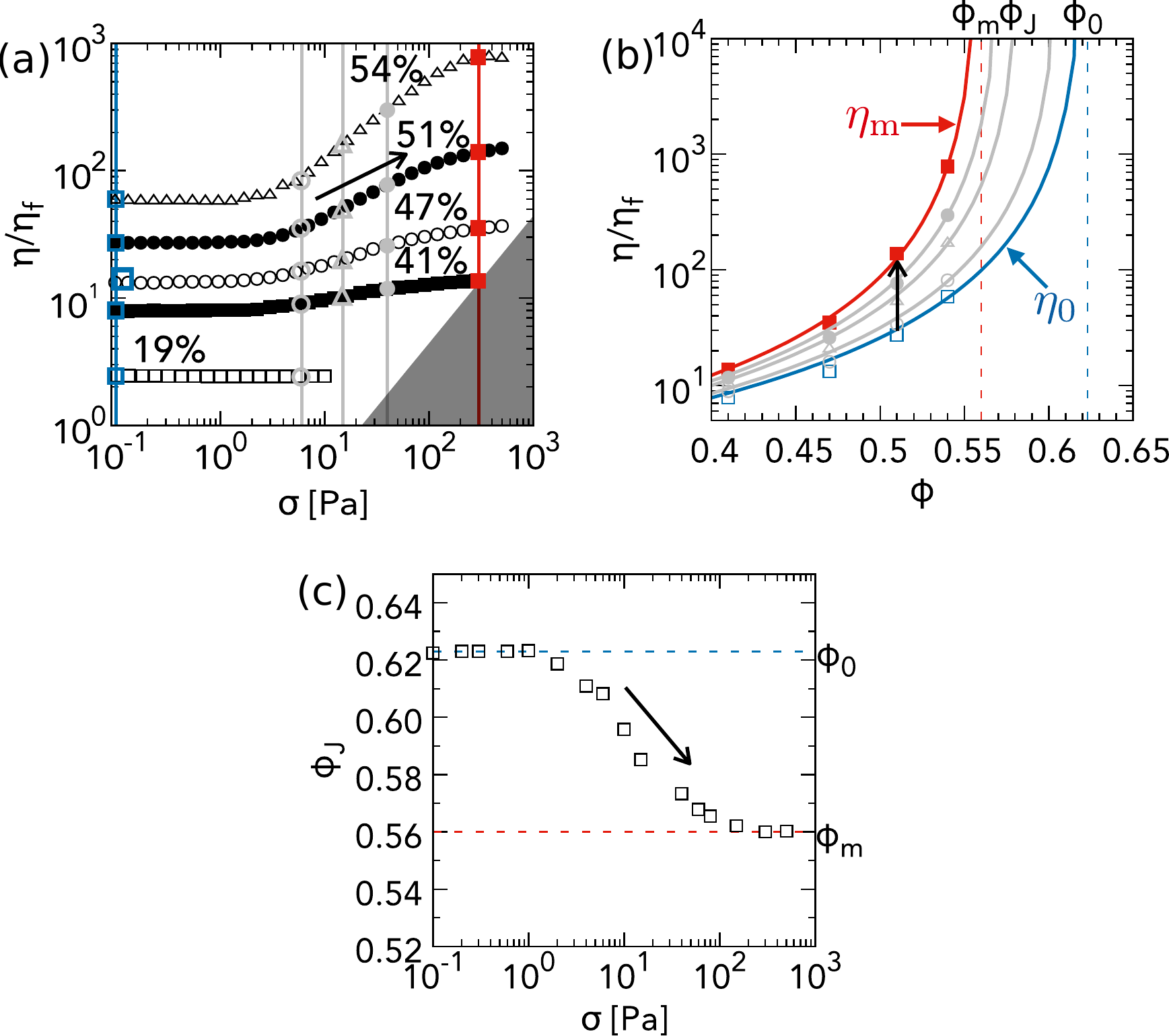}
\caption{\emph{Experimental shear thickening phenomenology.}
(a) Relative viscosity $\eta/\eta_f$ as a function of shear stress $\sigma$ at different volume fractions $\phi$ (as labelled) for $d=\SI{3.78}{\micro\metre}$, PHSA-stabilised, PMMA spheres in a cyclohexylbromide-decalin mixture of viscosity $\eta_f=\SI{2.83e-3}{\pascal.s}$ (taken from Guy~{\it et al.}\cite{guy2015towards}). The grey region is inaccessible due to inertial edge fracture.
(b) Symbols, viscosity ``branches" for different (fixed) values of $\sigma$ in (a). Lines correspond to fits by eye to Eq.~\ref{eq:div_simple}. The jamming volume fraction at which each viscosity branch diverges, $\phi_{\rm J}$, depends on~$\sigma$. Blue and red lines and symbols correspond respectively to the limiting low-$\sigma$ and high-$\sigma$ viscosities, $\eta_0$ and $\eta_{\rm m}$.
(c) $\phi_{\rm J}(\sigma)$, obtained from the fits shown in (b). $\phi_{\rm J}(\sigma)$ decreases smoothly from $\phi_0$, the $\phi$ at which $\eta_0$ diverges, to $\phi_{\rm m}$, the $\phi$ at which $\eta_{\rm m}$ diverges.
In all parts: shear thickening arises at any $\phi$, e.g., $\phi=0.51$, because increasing $\sigma$ decreases $\phi_{\rm J}$ [black arrow in (c)], shifting the viscosity branch in (b) to the left and so increasing $\eta$ [black arrows in (b) and (a)].
}
\label{fig:phenomenology}
\end{figure}

The phenomenology is generic\cite{guy2015towards}\cite{royer2016rheological}\cite{hodgson2015jamming}. Figure~\ref{fig:phenomenology}(a) shows  literature flow curves\cite{guy2015towards}, $\eta(\sigma)$, for buoyancy-matched suspensions of polymethylmethacrylate (PMMA) spheres with diameter $d \approx \SI{4}{\micro\metre}$ at different volume fractions $\phi$. At any fixed $\sigma$ (vertical lines), the viscosity increases with $\phi$, Fig.~\ref{fig:phenomenology}(b) (symbols). The viscosity ``branches'' at different $\sigma$ are well described by
\begin{equation}
\eta/\eta_f=(1-\phi/\phi_{\rm J})^{-2},
\label{eq:div_simple}
\end{equation}
which diverges at a $\sigma$-dependent jamming volume fraction $\phi_{\rm J}(\sigma)$; $\eta_f$ is the solvent viscosity.  Figure~\ref{fig:phenomenology}(b) shows example fits of Eq.~\ref{eq:div_simple} (lines) with $\phi_{\rm J}$ as a free parameter. The fitted $\phi_{\rm J}(\sigma)$, Fig.~\ref{fig:phenomenology}(c), is a \emph{decreasing} function of $\sigma$; so, increasing $\sigma$ at fixed $\phi$, i.e., traversing a vertical path in Fig.~\ref{fig:phenomenology}(b) (arrow), decreases $\phi_{\rm J}$ and causes $\eta$ to increase: the suspension shear thickens, Fig.~\ref{fig:phenomenology}(a). The limiting low- and high-$\sigma$ viscosity plateaux, $\eta_0$ and $\eta_{\rm m}$ [blue and red in Fig.~\ref{fig:phenomenology}(b)], diverge at $\phi_0$ and $\phi_{\rm m}<\phi_0$, respectively.

There is evidence of this scenario in a range of experimental systems\cite{guy2015towards,royer2016rheological,hodgson2015jamming}. The precise values of $\phi_0$ and $\phi_{\rm m}$, and the form of $\phi_{\rm J}(\sigma)$ [and hence $\eta(\sigma)$], depend on details of particle shape\cite{brown2010shear}, size polydispersity\cite{hodgson2015jamming} and surface roughness\cite{hsiao2017rheological,hsu2018roughness}. In all systems, shear-induced jamming\cite{peters2016direct}, inhomogeneous flow (shear banding)\cite{fall2015macroscopic} or unsteady flow\cite{hermes2016unsteady} are observed for $\phi_{\rm m} \leq \phi < \phi_0$, where conditions exist for which $\phi > \phi_{\rm J}$, Fig.~\ref{fig:phenomenology}(c), i.e., the system can exhibit solid-like behaviour.

\begin{figure}[t]
\centering
\includegraphics[width=\columnwidth]{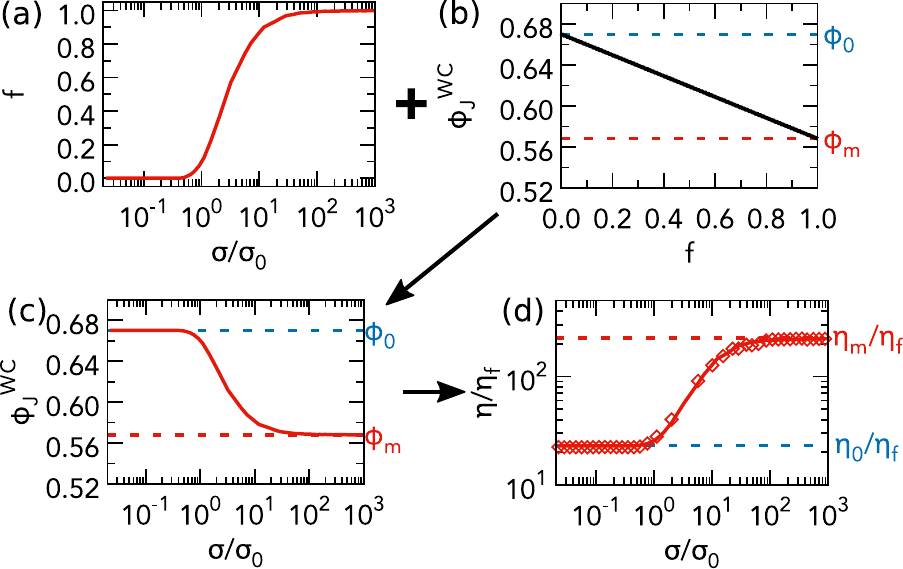}
\caption{ \emph{Logic of the WC shear thickening model.}
(a) The fraction of frictional contacts, $f$, takes a sigmoidal form. (b) $\phi_{\rm J}^{\rm WC}(f)$ is linearly interpolated between $\phi_0$ at $f = 0$ to $\phi_{\rm m}$ at $f = 1$. (c) The previous two plots directly give $\phi_{\rm J}^{\rm WC}(f(\sigma))$, which is inverse sigmoidal. (d) Using $\phi_{\rm J}^{\rm WC}(f(\sigma))$ in Eq.~\ref{eq:div_simple} gives $\eta(\sigma)$, which shows shear thickening (line). \emph{Testing the WC model using simulations.} The plot in (a) is calculated directly using contact forces from simulations of pure small-sphere supension at $\phi = 0.53$. See the text for how we obtain values for $\phi_0$ and $\rm \phi_m$ to calculate the $\phi_{\rm J}^{\rm WC}(f)$ plotted in (b) from Eq.~\ref{eq:phiJ}. These two plots directly give the $\phi_{\rm J}^{\rm WC}$ in (c), which, when used in Eq.~\ref{eq:div_simple} gives the flow curve in (d), $\eta^{\rm WC}(\sigma)$ (line). The symbols in (d) are the computed viscosity from simulations.
}
\label{fig:WC}
\end{figure}

In the WC model\cite{wyart2014discontinuous}, inter-particle contacts are either lubricated, with static friction coefficient $\mu=0$, or frictional, with $\mu>0$. The fraction of the latter, $f$, increases with $\sigma$, Fig.~\ref{fig:WC}(a). WC's jamming volume fraction is a function of $f$ only:
\begin{equation}
\phi_{\rm J}^{\rm WC}=f\phi_{\rm m} + (1-f)\phi_0,
\label{eq:phiJ}
\end{equation}
changing linearly from random close packing, $\phi_0$, at $f=0$ (all lubricated contacts) to frictional jamming, $\phi_{\rm m}$, at $f=1$ (all frictional contacts), Fig.~\ref{fig:WC}(b). Thus, $\phi_{\rm J}^{\rm WC}(f(\sigma))$ decreases with $\sigma$, Fig.~\ref{fig:WC}(c), and determines $\eta$ via some empirical form, e.g., Eq.~\ref{eq:div_simple}\cite{guy2015towards,royer2016rheological}, leading to shear-thickening flow curves, Fig.~\ref{fig:WC}(d) (line).

The WC model, Eq.~\ref{eq:div_simple}-\ref{eq:phiJ}, predicts the $\sigma$- and $\phi$- dependent viscosity, $\eta^{\rm WC}(\sigma,\phi)$, from three inputs: the limiting frictionless and frictional jamming points, $\phi_0$ and $\phi_{\rm m}$, and the $\sigma$-dependent fraction of frictional contacts, $f$. $\phi_{\rm m}$ and $\phi_0$, can be obtained by fitting viscosity branches at different $\sigma$, as done in Fig.~\ref{fig:phenomenology}(b). They are not related to shear thickening \emph{per se}. On the other hand, $f$, which determines the shape of the flow curve, is currently inaccessible in experiments. 
Thus, various {\it ansatzs} are used to fit the WC model to experiments. For sterically-stabilised PMMA spheres, Guy \emph{et~al.}\cite{guy2015towards} used a $\phi$-independent sigmoidal form:
\begin{equation}
    f(\sigma)=\exp[-(\sigma^*/\sigma)^{\beta}],
    \label{eq:guy}
\end{equation}
with $\beta=0.85$. The single stress scale, $\sigma^*$, scales as the ``engineering'' onset stress at which $\eta(\sigma)$ visibly begins to increase.  

Importantly for this study, we note that the particle size does {\it not} appear in WC model. On the other hand, and perhaps significantly in light of the findings we will report, Guy \emph{et~al.} found that the onset stress decreases with particle size, with $\sigma^* \propto d^{-2}$, suggesting $\sigma^* \propto F^*/d^2$ for their PMMA particles, where $F^*\sim k_BT/\SI{}{\nano\metre}$ is a constant force\cite{wyart2014discontinuous,guy2015towards}. Royer \emph{et al.}\cite{royer2016rheological} used a similar form to fit data for dispersions of charge-stabilised silica.

In discrete-element (DEM) simulations, $f$ can be calculated directly from particle coordinates. A popular choice is to use the ``critical-load model'' (CLM), in which $\mu$ jumps from 0 to $>0$ when the normal contact force between particles, $F$, exceeds a threshold value, $F^*$ (the critical load). This model reproduces\cite{mari2014shear} the phenomenology of Fig.~\ref{fig:phenomenology} and unstable flow at high $\phi$. 
For a bidisperse mixture of spheres with diameter $d_1$ and $d_2=d_1/1.4$, Mari \emph{et~al.}~found\cite{mari2014shear} a $\phi$-independent $f(\sigma)$ of the form Eq.~\ref{eq:guy},
with $\beta=1.1$ and $\sigma^* \approx F^*/[6\pi (d_2/2)^2]$, and  later used it to fit flow curves at a range of $\phi$\cite{singh2018constitutive}. Thus, in this one case, the WC model is fully validated: using the measured fraction of frictional contacts in Eq.~\ref{eq:phiJ} correctly predicts the viscosity. The similarity between the forms of $f(\sigma)$ used to fit experiments and measured in simulations suggests that $f$ in mildly polydisperse experimental systems can indeed be well described by Eq.~\ref{eq:guy} or some similar form.

In these studies, a weak polydispersity (= standard deviation normalised by the mean of the particle size distribution) of $s \lesssim 20\%$ was used to inhibit shear-induced crystallisation\cite{mari2014shear}. Industrial dispersions typically have broad, often multimodal, size distributions with  $s \gtrsim 100\%$. Nevertheless, the low-$s$ phenomenology in Fig.~\ref{fig:phenomenology} continues to hold\cite{d1994rheological,bender1996reversible,hodgson2015jamming,guy2017physics}. However, the validity of the WC model in such higher-$s$ suspensions has not been tested.

Indeed, it is a surprise for the WC model to work, and work well, even for low-$s$ systems. In suspension rheology,  details of the microstructure, fabric of the contact network and distribution of forces matter. Cates pointed out long ago that the relatively small number of nearest neighbours, $\sim\mathcal{O}(10)$, usually precludes any mean-field description\cite{Cates2003}. The success of his shear-thickening  model with Wyart contradicts this norm. In the WC model, $\eta(\sigma)$ is controlled primarily by a single scalar parameter $f$ that is agnostic to exact microstructural details. For this reason alone, it is important to test the limitations of the WC model.

Here, we do so in suspensions of strongly bidisperse spheres. 
As before\cite{mari2014shear,singh2018constitutive}, we use DEM simulations to extract $f(\sigma)$ for different mixtures and compare the predictions of the WC model to bulk flow curves. We then use the same $f(\sigma)$ to test the model against experimental data for bidisperse PMMA spheres. We find that WC works for nearly-monodisperse suspensions [i.e., the simulated $f(\sigma)$ correctly predicts $\eta(\sigma)$], but fails in general for bidisperse suspensions. We show that, nevertheless, the model can be extended to fit our data if the fractions of each contact type (large-large, large-small and small-small) are taken into account separately. Our results indicate that, in its original form, the WC model is at least incomplete, and highlight a number of unresolved issues in the current understanding of shear thickening and in the use of the WC framework to make inferences about microphysics. We propose directions for future research to address these issues.

\section{Methods}
A binary mixture of spheres is characterised by four parameters: $d_1$, $d_2<d_1$, the fraction of small particles $\xi=V_2/(V_1+V_2)$ (where $V_1$ and $V_2$ are the total volumes of large and small particles, respectively) and the total volume fraction $\phi=(V_1+V_2)/V$ (where $V$ is the total volume). We fix $\phi$ and the size ratio $\alpha\equiv d_2/d_1 \approx 0.25$, and vary $\xi$. At our $\alpha$, the small spheres are slightly too large to fit inside four touching large spheres (for which $\alpha_{\rm max} = 0.225$). 

We sheared $N=2000$ bidisperse, repulsive spheres at fixed $\phi$ in a periodic cell with Lees-Edwards boundary conditions. Short-range lubrication and repulsive contact forces described by linear springs of stiffness $k$ were resolved using a classical DEM code that allows marginal overlaps $\delta$ between the surfaces of pairs of particles\cite{ness2016shear}. We employed a contact model\cite{mari2014shear} in which Coulomb friction with static friction coefficient $\mu=1$ appears beyond a critical overlap $\delta^*$, corresponding to a critical normal load $F^*=k \delta^*$.
For simplicity, and consistency with experiments for nearly-monodisperse systems\cite{guy2015towards}, we use a constant $F^*$ (and hence $\delta^*$) that is independent of $d_1$ and $d_2$.

Our unit of stress is $\sigma_0=F^*/(3\pi d_2^2/2)$, at which purely small particles ($\xi=1$) shear thicken\cite{mari2014shear}; pure large spheres shear thicken at $\sim \alpha^2 \sigma_0$. For bidisperse mixtures, we averaged $\sigma$ over the strain interval $\gamma \subset [1.5,3]$ or $\subset [1.5,2]$, in which the system had reached steady state for all $\xi$. For monodisperse suspensions,  we averaged over $\gamma \subset [0.7,1]$ to avoid the onset of large-scale crystallization\cite{sierou2002rheology}.

We performed experiments on binary suspensions of PMMA spheres stabilised with poly-12-hydroxystearic acid (PHSA) in a near-density-matching mixture of cyclohexylbromide and decalin (density $\approx$\SI{1.18}{g.cm^{-3}}, viscosity $\eta_f=$\SI{2.4}{\milli Pa.s}). 
We varied $\xi$ at a fixed $\phi=0.51$ by mixing together monomodal suspensions with mean particle diameters $d_2=\SI{0.712}{\micro\metre}$ and $d_1=\SI{2.76}{\micro\metre}$ and $s \approx 10\%$ (from static light scattering). Flow curves were measured using an Anton Paar MCR 301 rheometer with sandblasted steel cone (angle 1$^{\circ}$; diameter \SI{50}{\milli \metre}; truncation \SI{100}{\micro\metre}; roughness $\sim$\SI{10}{\micro \metre}) and roughened aluminium base plate (roughness $\sim$\SI{10}{\micro \metre}) at \SI{20}{\celsius}. A solvent trap minimised evaporation. Details of experiments are given in the Electronic Supplementary Information (ESI)$^{\dag}$.

\section{Results}

\subsection{Bidisperse shear thickening phenomenology}
\begin{figure}[t]
\centering
\begin{minipage}{0.49\columnwidth}
\includegraphics[width=\textwidth]{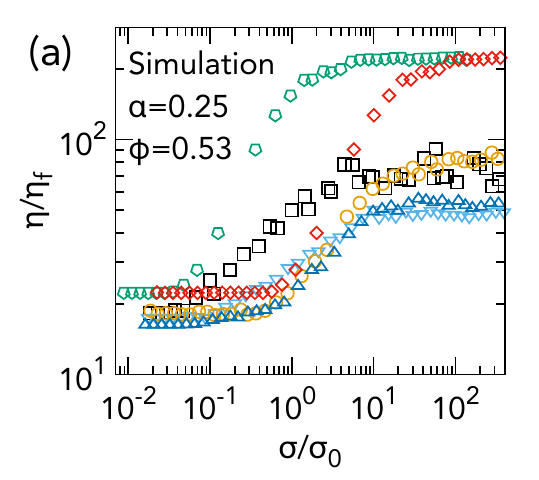}
\end{minipage}
\begin{minipage}{0.49\columnwidth}
\includegraphics[width=\textwidth]{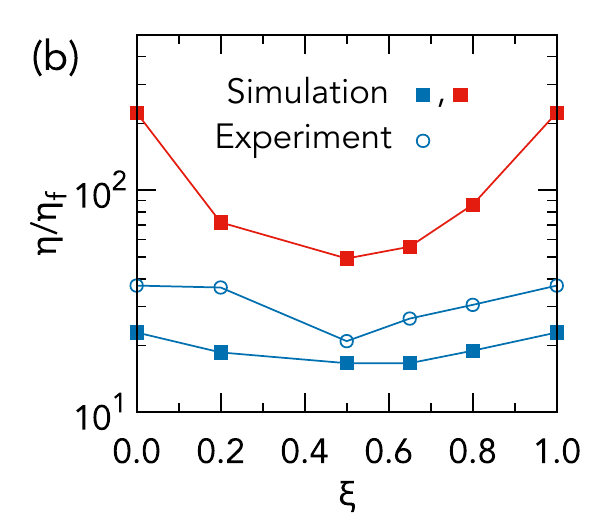}
\end{minipage}
\\
\begin{minipage}{0.49\columnwidth}
\includegraphics[width=\textwidth]{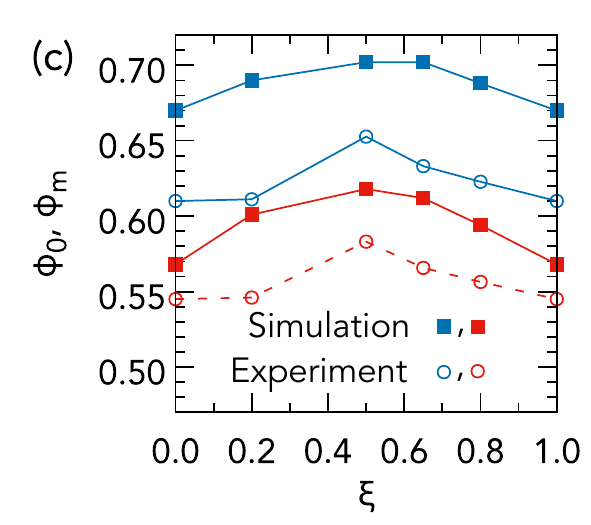}
\end{minipage}
\begin{minipage}{0.49\columnwidth}
    \includegraphics[width=1\textwidth]{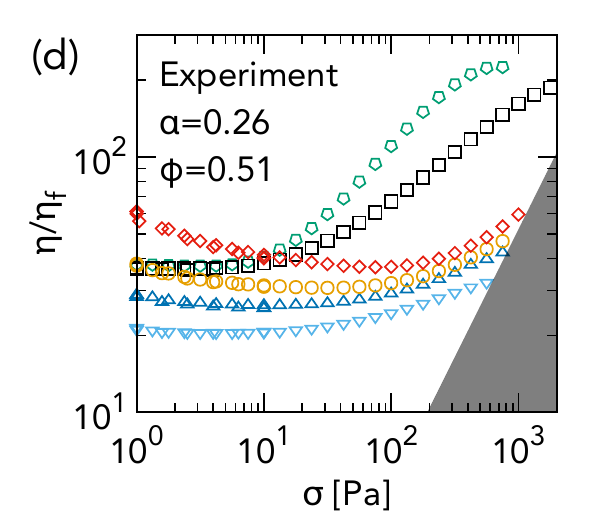}
\end{minipage}
\caption{\emph{Bidisperse shear thickening phenomenology.}
(a) $\eta/\eta_f$ as a function of $\sigma/\sigma_0$ from simulations at $\alpha=0.25$ and $\phi=0.53$, with $\xi=0$ (pentagon), 0.2 ($\square$), 0.5 ($\triangledown$), 0.65 ($\triangle$), 0.8 ($\circ$) and 1 ($\diamond$). 
(b) Frictionless relative viscosity $\eta_0/\eta_f$ from simulations (\blue{--$\blacksquare$--}) and experiments (\blue{--$\circ$--}), and frictional relative viscosity $\eta_{\rm m}/\eta_f$ from simulations (\red{--$\blacksquare$--}). (c) Limiting jamming volume fractions, $\phi_0$ (blue) and $\phi_{\rm m}<\phi_0$ (red), versus $\xi$ from simulations (\blue{--$\blacksquare$--},\red{--$\blacksquare$--}) and experiments (\blue{--$\circ$--},\red{--$\circ$--}).
(d) Experimental $\eta/\eta_f$ versus $\sigma$ for PMMA spheres at $\alpha=0.26$ and $\phi=0.51$, with $\xi=0$ (pentagon), 0.2 ($\square$), 0.5 ($\triangledown$), 0.65 ($\triangle$), 0.8 ($\circ$) and 1 ($\diamond$). Inertial fracture at $\dot{\gamma} \approx \SI{8e3}{s^{-1}}$ renders the grey-shaded region inaccessible\cite{guy2015towards}.
}
\label{fig:sim}
\end{figure}

We first present the simulated relative viscosity $\eta/\eta_f$ as a function of shear stress $\sigma/\sigma_0$ for fixed $\phi=0.53$ and $\alpha=0.25$ at various fractions of small particles, Fig.~\ref{fig:sim}(a), $\xi=0$ (pentagon), 0.2 ($\square$), 0.5 ($\triangledown$), 0.65 ($\triangle$), 0.8 ($\circ$) and 1 ($\diamond$)\footnote{By \emph{number}, small particles dominate the large particles for all the $\xi$ we study. For the data in Fig.~\ref{fig:sim}(a), the number fractions of small particles, $x=1/[1+\alpha^3(1/\xi-1)]$, are: $x=0$ (pentagon), 0.941 ($\square$), 0.985 ($\triangledown$), 0.992 ($\triangle$), 0.996 ($\circ$) and 1 ($\diamond$).}.

Bidisperse and monodisperse flow curves are qualitatively similar, showing shear thickening between two Newtonian plateaux. Figure~\ref{fig:sim}(b) shows the $\xi$-dependent plateau viscosities, $\eta_0(\xi)$ (\blue{$\blacksquare$}) and $\eta_{\rm m}(\xi)$ (\red{$\blacksquare$}), estimated by eye from Fig.~\ref{fig:sim}(a). Mixing particles reduces both limiting viscosities relative to the values for single-sized spheres. Such a ``Farris effect''\cite{farris1968prediction}  has been widely observed in fixed-friction (i.e., Newtonian) suspensions\cite{pednekar2018bidisperse,farris1968prediction}. The limiting volume fractions, $\phi_0(\xi)$ and $\phi_{\rm m}(\xi)$, Fig.~\ref{fig:sim}(c) (\blue{$\blacksquare$} and \red{$\blacksquare$} respectively), are calculated using the simulated $\eta_0$ and $\eta_{\rm m}$ values in Eq.~\ref{eq:div_simple} with $\phi = 0.53$. The non-monotonic behaviour directly mirrors that of $\eta_0(\xi)$ and $\eta_{\rm m}(\xi)$.

Experimental flow curves for binary mixtures of PMMA with $\alpha=0.26$ and $\phi=0.51$, Fig.~\ref{fig:sim}(d), show similar phenomenology, except that  the limiting high-$\sigma$ behaviour is preempted by edge fracture due either to an inertial instability (grey region), or a different fracture mechanism when $\sigma$ exceeds $\approx 10^3~\si{\pascal}$$^{\dag}$. Thus, we cannot access $\eta_{\rm m}(\xi)$ directly for all $\xi$. Shear thickening is preceded by shear thinning, presumably due to residual Brownian motion\cite{guy2015towards}\footnote{Thus, the viscosity of the small spheres is greater than that of the large spheres below the onset of thickening, e.g., at $\sigma=\SI{1}{\pascal}$.}; so, we estimate $\eta_0$ by  the viscosity minimum before the onset of thickening, Fig.~\ref{fig:sim}(b) (\blue{$\circ$}). The experimental $\eta_0(\xi)$ show the same non-monotonicity as the simulated values, but are always too high, by up to a factor of $\lesssim 2$ for the two end members ($\xi = 0$ or 1). Using the experimental values of $\eta_0(\xi)$ in Eq.~\ref{eq:div_simple} with $\phi = 0.53$ gives us an experimental estimate of $\phi_0(\xi)$, Fig.~\ref{fig:sim}(c) (\blue{$\circ$}). Consistent with the experimental viscosities $\eta_0(\xi)$ being higher than simulated values, the experimentally deduced $\phi_0(\xi)$ are somewhat lower than the simulated values, i.e.~the experimental system at $\phi = 0.53$ is closer to jamming than the corresponding simulated system.

\subsection{Comparing simulations to the WC model}

We test the WC model by comparing simulated flow curves, $\eta(\sigma)$, with those calculated using the fraction of frictional contacts, $f(\sigma)$, measured from the simulations, $\eta^{\rm WC}(\sigma)$. 
To explain our procedure, consider data for monodisperse small particles ($\xi = 1$).
First, we calculate $f(\sigma)$ directly from inter-particle forces by counting, at each $\sigma$, the fraction of contacts with $F>F^*$. The $f(\sigma)$ so obtained, Fig.~\ref{fig:WC}(a), is sigmoidal, similar to the $f(\sigma)$ in bidisperse mixtures\cite{mari2014shear} with $\alpha=0.71$ and Eq.~\ref{eq:guy}. 

To calculate $\phi_{\rm J}^{\rm WC}(f)$ from Eq.~\ref{eq:phiJ}, we need $\phi_0$ and $\phi_{\rm m}$, which could be obtained by simulating and fitting $\eta(\sigma,\phi)$ at a range of $\phi$, as done in Fig.~\ref{fig:phenomenology}. Instead, we use our simulated values of the low- and high-$\sigma$ viscosities, $\eta_0$ and $\eta_{\rm m}$, at $\phi = 0.53$ in Eq.~\ref{eq:div_simple} to obtain $\phi_0$ and $\phi_{\rm m}$, giving the $\phi_{\rm J}^{\rm WC}(f)$ in Fig.~\ref{fig:WC}(b).\footnote{We expect this to be a reasonable approximation, since Eq.~\ref{eq:div_simple} has previously been used to fit $\eta(\phi)$ for various frictional, bidisperse sphere mixtures\cite{pednekar2018bidisperse}.}

From $f(\sigma)$, Fig.~\ref{fig:WC}(a), and $\phi_{\rm J}^{\rm WC}(f)$, Fig.~\ref{fig:WC}(b), we now calculate  $\phi_{\rm J}^{\rm WC}(f(\sigma))$, Fig.~\ref{fig:WC}(c), which decreases smoothly from $\approx\phi_0$ at $\sigma/\sigma_0 \ll 1$ to $\phi_{\rm m}$ at $\sigma/\sigma_0 \gg 1$.
Finally, we calculate the viscosity by substituting $\phi_{\rm J}^{\rm WC}(f(\sigma))$, Fig.~\ref{fig:WC}(c), into Eq.~\ref{eq:div_simple}. The flow curve, $\eta^{\rm WC}(\sigma)$, Fig.~\ref{fig:WC}(d) (solid line), increases smoothly from $\eta_0$ to $\eta_{\rm m}$.

\begin{figure}[h!]
\centering
\begin{minipage}{0.49\columnwidth}
    \includegraphics[width=1.0\textwidth]{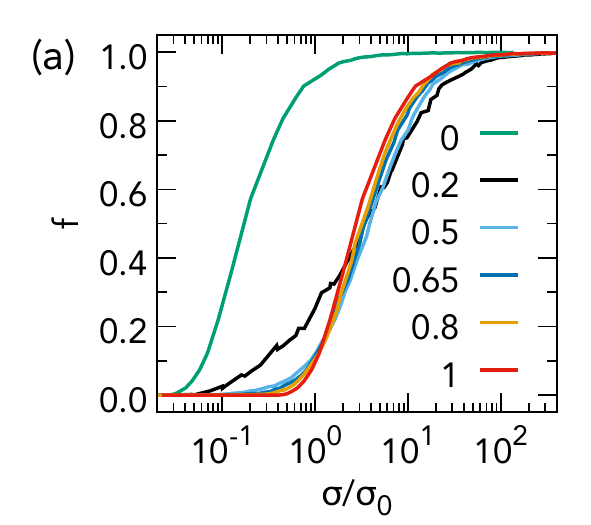}
\end{minipage}
\begin{minipage}{0.49\columnwidth}
  \includegraphics[width=1.0\textwidth]{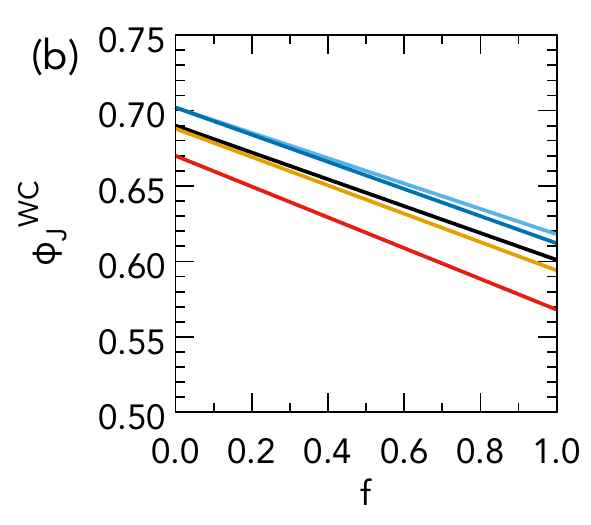}
\end{minipage}
\\
\begin{minipage}{0.49\columnwidth}
  \includegraphics[width=1.0\textwidth]{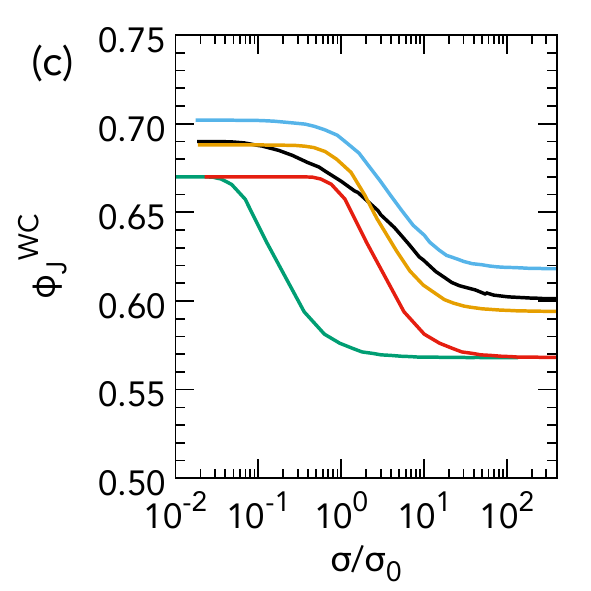}
\end{minipage}
\begin{minipage}{0.49\columnwidth}
  \includegraphics[width=1.0\textwidth]{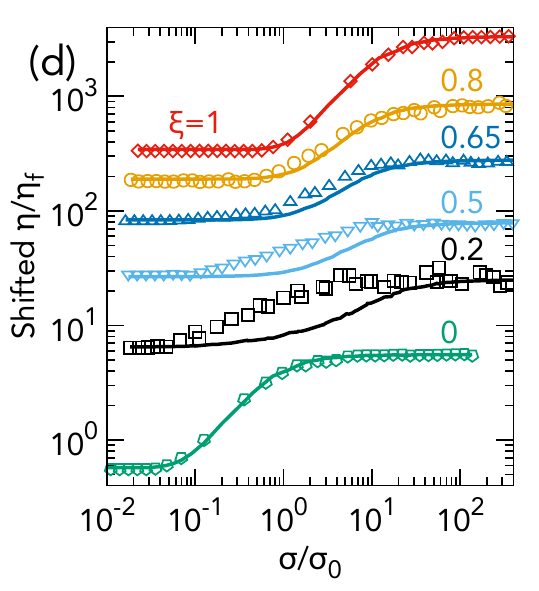}
\end{minipage}
\caption{\emph{Failure of the WC model for bidisperse simulations.}
(a) Fraction of frictional contacts $f$ as a function of $\sigma/\sigma_0$, extracted from simulations at different $\xi$, as labelled.
(b) WC jamming volume fraction, $\phi_{\rm J}^{\rm WC}$, as a function of $f$ for different $\xi$ [colours as in part (a)].
(c) $\phi_{\rm J}^{\rm WC}$ as a function of $\sigma$ calculated using (b) and $f(\sigma)$ from (a).
(d) Symbols: shifted flow curves for different $\xi$, as labelled. Shift factors are: $\xi=0$, 0.025; $\xi=0.2$, 0.35; $\xi=0.5$, 1.6; $\xi=0.65$, 5; $\xi=0.8$, 10 and $\xi=1$, 15. Lines: predictions of the WC model shifted by the same factors.
}
\label{fig:WC_fails}
\end{figure}

We compare this flow curve calculated using the measured $f(\sigma)$ with the simulated $\eta(\sigma)$, Fig.~\ref{fig:WC}(d) (symbols). The two calculated plateaux agree with the simulated values by construction. The WC model is judged instead by how well it captures the shear thickening process. It does this well for monodisperse spheres. Both model-predicted and simulation viscosities start to increase at $\sigma/\sigma_0 \approx 1$, reaching saturation at $\sigma/\sigma_0 \gtrsim 30$. 

We repeat this procedure for our bidisperse suspensions with $\xi=0.2,0.5,0.65$ and $0.8$. Again, the measured $f(\sigma)$, Fig.~\ref{fig:WC_fails}(a), and linearly interpolated $\phi_{\rm J}^{\rm WC}(f)$, Fig.~\ref{fig:WC_fails}(b), are used to calculate $\phi_{\rm J}^{\rm WC}(\sigma)$, Fig.~\ref{fig:WC_fails}(c), from which we obtain flow curves, Fig.~\ref{fig:WC_fails}(d) (lines). We compare these with the simulated viscosities, Fig.~\ref{fig:WC_fails}(d) (symbols), recalling that the limiting viscosities are constrained to fit, and noting that data for different $\xi$ have been shifted vertically to aid visibility (see caption for shift factors). 

Note, first, that $f(\sigma)$ for the monodisperse end members, $\xi = 0$ and 1, are identical in shape, Fig.~\ref{fig:WC_fails}(a), but with the former shifted to the left by a factor of $(d_2/d_1)^2=\alpha^2=0.0625$ due to a decrease in $\sigma^*$ by the same factor for the larger particles\cite{guy2015towards}. Addition of just 20\% of small spheres to a suspension of large spheres ($\xi = 0.2$) produces a dramatic effect, Fig.~\ref{fig:WC_fails}(a). While frictional contacts still start to form at $\sigma^*_1 \approx 0.06\sigma_0$, the onset is now much more gradual, until $\sigma \approx \sigma^*_2 \approx \sigma_0$, whereupon ${\rm d}f/{\rm d} \sigma$ abruptly becomes as large as the monodisperse case (either $\xi = 0$ or 1), before $f$ saturates at a $\sigma$ that is comparable to (but slightly later than) that of monodisperse small spheres, even though only 20\% of these are present. 
By $\xi = 0.5$, $f(\sigma)$ becomes very similar to that of that of monodisperse small spheres ($\xi = 1$); although, the onset is still clearly somewhat earlier and the saturation somewhat lower. These features become progressively less obvious at $\xi = 0.65$ and 0.80. The effect of bidispersity is therefore asymmetrical: the presence of 20\% of large spheres in 80\% of small spheres has far smaller an effect on $f(\sigma)$ than the reverse situation. 

Turning to the shear-thickening flow curves, Fig.~\ref{fig:WC_fails}(d), we see that, as expected, the WC model reproduces the simulated data for the two monodisperse end members. It gives a tolerable representation of the data at $\xi = 0.8$, i.e. when there are 20\% of large spheres present in a predominantly small-sphere system; but, it fails badly in the reverse situation, when there are 20\% of small spheres in a mainly large-sphere system ($\xi = 0.2$). The disagreement between the WC prediction and simulation data is still substantial at $\xi = 0.5$, and remains perceptible at $\xi = 0.65$.

\subsection{Comparing experiments to the WC model} 
Testing the WC model against experimental data is more involved. Figure~\ref{fig:sim}(a) shows clearly that introducing bidispersity alters shear thickening; however,  simulations\cite{mari2014shear} and recent experiments\cite{comtet2017pairwise,chatte2018shear} have shown that, even for nearly monodisperse suspensions, thickening is also sensitive to the relationship between the particle static friction coefficient, $\mu$, and the normal contact force, $F$.  The function $\mu(F)$ is fully prescribed in our simulations: $\mu=0$ below a threshold force $F^*$, and $\mu=\mu_{\rm m}>0$ above $F^*$, allowing us to isolate the effect of bidispersity on the shear-thickening phenomenology. However, for our sterically stabilised PMMA particles the load-dependent friction $\mu(F)$ has not been measured; hence,  we do not know \emph{a priori} the role of the specific tribology of our particles.

For simplicity, we assume that the experimental $\mu(F)$ obeys the same contact model (CLM) as in simulations and treat the critical load, $F^*$, as an unknown parameter; as in simulations, we take $F^*$ to be independent of the size of the contacting particles.  $F^*$ defines a contact stress scale $\sigma_0 \sim F^*/d_2^2$, the unit of stress in our simulations. For each $\sigma$ and $\xi$ in experiments we calculate $\sigma/\sigma_0$ and read off the corresponding fraction of frictional contacts, $f(\sigma/\sigma_0)$, from simulations, Fig.~\ref{fig:WC_fails}(a).

\begin{figure}[t]
\centering
\begin{minipage}{0.33\textwidth}
\includegraphics[width=1.0\textwidth]{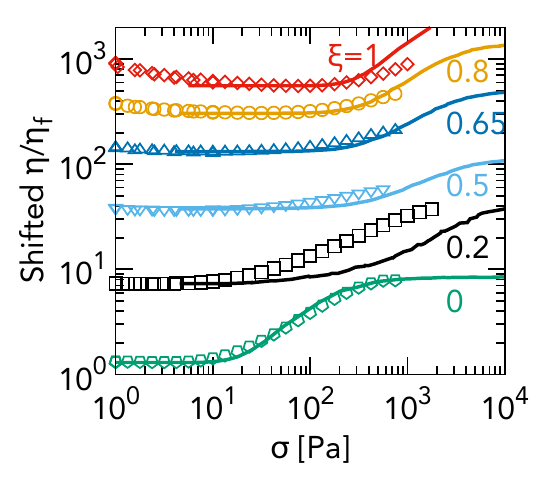}
\end{minipage}
\caption{\emph{Failure of the WC model for PMMA spheres.}
Symbols, shifted flow curves from experiments for different $\xi$, as labelled. Lines, shifted WC model predictions. Shift factors are: $\xi=0$, 0.035; $\xi=0.2$, 0.2; $\xi=0.5$, 1.8; $\xi=0.65$, 5; $\xi=0.8$, 10 and $\xi=1$, 15.
}
\label{fig:failure_experiment}
\end{figure}

Using $\sigma_0$ as a global fitting parameter, and $\phi_{\rm m}$ and $\phi_0$ as local fitting parameters, we use the $f(\sigma/\sigma_0)$ so obtained to calculate $\phi_{\rm J}(\sigma)$ via Eq.~\ref{eq:phiJ}, from which we compute $\eta^{\rm WC}(\sigma)$ with Eq.~\ref{eq:div_simple}. In Fig.~\ref{fig:failure_experiment}, we plot measured flow curves (symbols) and WC flow curves (lines) for $\sigma_0=\SI{250}{\pascal}$ and $\phi_{\rm m}(\xi)=\Lambda \phi_0(\xi)$ with $\Lambda=0.89$. Choosing a $\xi$-dependent $\Lambda=\phi_{\rm m}/\phi_0$ does not affect any of our conclusions. Data and fits have been shifted vertically for clarity (see caption).

The all-large ($\xi=0$) flow curve is well fit by the WC model, in agreement with our simulations. This  justifies \emph{a posteriori} our use of the CLM for $\mu(F)$ for this sample. The model should equally well describe the all-small ($\xi=1$) flow curve; however, this is not the case. Although the onset of shear thickening is correctly predicted, the rise in $\eta(\sigma)$ is overestimated by the model, implying a different $\mu(F)$ for the small particles, e.g., CLM with a lower $\mu_{\rm m}$ than for the large particles\cite{mari2014shear}. As a consequence, the present map between $\sigma$ and $f$ is not reliable for our bidisperse suspensions; to calculate $f(\sigma)$ properly, one would have to independently characterise $\mu(F)$ experimentally for the different contact types (large-large, small-small and large-small) and compare with simulations employing a representative DEM contact model. We do not do so here; however, we point out that all the same trends noted when we compared simulations with the WC model are clearly reproduced in our bidisperse experiments ([$0<\xi<1$ in Fig.~\ref{fig:failure_experiment}]). In particular, there is a striking disagreement between model and experiment at $\xi=0.2$, for which, as in Fig.~\ref{fig:WC_fails}(d), the onset of shear thickening is grossly underestimated. Based on this similarity, we can infer already that the same shortcomings of the WC model as applied to bidisperse simulations should also apply to bidisperse experiments.

\section{How the WC model fails}
Previous experiments and simulations find, and we confirm, that the WC model works well in the quasi-monodisperse limit ($s \lesssim 0.2$)\cite{singh2018constitutive}. This phenomenological model is designed to reveal the consequences of a simplified set of assumptions in the most perspicuous way, and (according to its authors\cite{wyart2014discontinuous}) not meant for the fitting of data. Thus, that it works quantitatively in the small-$s$ limit is itself non-trivial, especially given its mean-field nature\cite{Cates2003}. 

It is perhaps unsurprising that we find the WC model fails to account for a binary mixture with $\alpha = 0.25$ (size ratio 1:4), which translates, using a previously-proposed definition\cite{voivret2009multiscale}, to an effective polydispersity $s_{\rm eff} = (d_1 - d_2)/(d_1 + d_2) \approx 60\%$ \footnote{A more natural definition would normalise to the average size to give $s \approx 120\%$.}. The pertinent question is: precisely where is the WC model failing in this case?

\begin{figure*}[t]
\centering
\includegraphics[width=\textwidth]{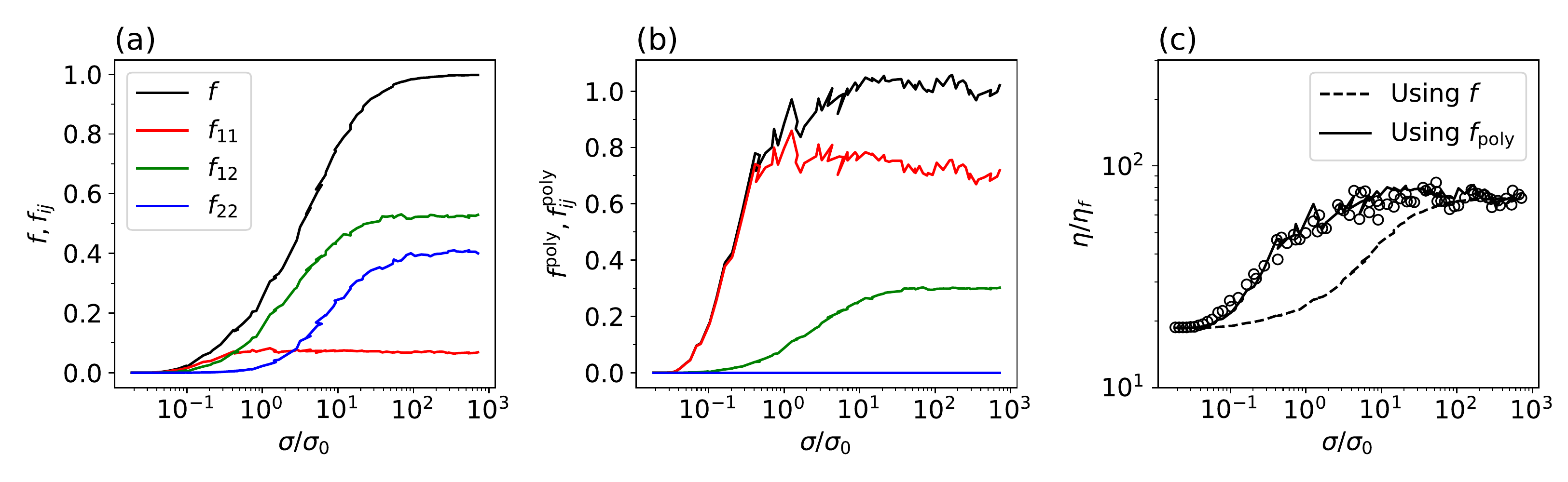}
\caption{\emph{WC model fails due to equal weighting of contact types}.
(a) Simulated $\sigma$-dependent fraction of frictional contacts, $f(\sigma)$ assumed implicitly by WC in their original model (black line), for $\xi=0.2$ decomposed into contributions from large-large, $f_{11}$; large-small, $f_{12}$ and small-small, $f_{22}$, contacts, as labelled.
(b) Weighted total fraction of frictional contacts $f_{\rm poly}$ needed to fit the WC model to our data, Eq.~\ref{eq:newphiJ} (black line), and individual weighted fractions, $f_{11}^{\rm poly}$ (red), $f_{12}^{\rm poly}$ (green) and $f_{22}^{\rm poly}$ (blue), as defined in Eq.~\ref{eq:fpoly}.
(c) Symbols, simulation flow curve. Dashed line, prediction of the WC model using $f$. Solid line, prediction of the WC model using $f^{\rm poly}$.
}
\label{fig:fij}
\end{figure*}

In a monodisperse system, there is a single kind of frictional contact. In a bidisperse system, such contacts are of three kinds: large-large (`11'), large-small (`12'), and small-small (`22'). Figure~\ref{fig:fij}(a) shows how the three types of frictional contact develop with stress, $f_{11}(\sigma)$, $f_{12}(\sigma)$ and $f_{22}(\sigma)$, in our simulated $\xi = 0.2$ system, where we observe maximal discrepancy  with the WC model. Not surprisingly, frictional contacts first form amongst the large species, at $\sigma^*_1 \approx 0.06\sigma_0$; however, this contribution rapidly saturates to $f_{11}^\infty \lesssim 0.1$. The latest frictional contacts to form are the small-small ones: $f_{22}$ does not start to increase until $\sigma^*_2 \approx \sigma_0$; but, these saturate to about $f_{22}^\infty \approx 0.4 \gtrsim 4f_{11}^\infty$. Ultimately, the largest contribution is from `mixed' contacts, $f_{12}^\infty \approx 0.5 \gtrsim 5f_{11}^\infty$, which start to form at $\sigma^*_{12} \approx 0.2\sigma_0$ (perhaps fortuitously close to $\sqrt{\sigma^*_1\sigma^*_2}$).

In its original form, the WC model is agnostic to particle size. Consistency with this feature requires that, when applied to our biphasic system, we take $f = f_{11}+f_{12}+f_{22}$, so that {\it any} new frictional contact formed as stress builds up contributes equally to the lowering of $\phi_{\rm J}^{\rm WC}$, and therefore to the viscosity increment via Eqs.~\ref{eq:div_simple} and \ref{eq:phiJ}. Thus, because $f_{12} \approx f_{22} \gg f_{11}$ at $\xi = 0.2$, the WC flow curve at this composition  is much more similar in shape to that for the all-small ($\xi = 1$) system than that for the $\xi = 0$ system. In reality, the simulated flow curves start to shear thicken at $\sigma^*_1$, which is where $f_{11}$ starts to increase; i.e., large-large contacts dominate despite the smallness of $f_{11}$, and  many small-large and small-small contacts seem to contribute little to the shift in $\phi_{\rm J}$.

This suggests that we should write $\phi_{\rm J}$ a function of $f_{11}$, $f_{12}$ and $f_{22}$, separately. A simple \emph{ansatz} is to retain the functional form of $\phi_{\rm J}$ in Eq.~\ref{eq:phiJ} and replace $f$ with a polydisperse crossover function, $f^{\rm poly}=f_{11}^{\rm poly}+f_{12}^{\rm poly}+f_{22}^{\rm poly}$, where the weighted fraction of frictional contacts for contacts of type $(ij)$ is $f^{\rm poly}_{ij}=\kappa_{ij}/f_{ij}^{\infty}$. The coefficient $\kappa_{ij}$ corresponds to the large-$\sigma$ limit of $f_{ij}^{\rm poly}$; $f_{ij}^\infty$ denotes the corresponding large-$\sigma$ limit of $f_{ij}$ in Fig~\ref{fig:fij}(a). We choose $\kappa_{11}+\kappa_{12}+\kappa_{22}=1$ to ensure  $f_{\rm poly}(\sigma/\sigma_0 \ll 1)=0$ and $f_{\rm poly}(\sigma/\sigma_0 \gg 1)=1$. So, our extended WC model reads
\begin{eqnarray}
	\phi_{\rm J}=f^{\rm poly}\phi_{\rm m} + (1-f^{\rm poly})\phi_0,\;\; \;\;\mbox{with}
	\label{eq:newphiJ}\\
f_{\rm poly}= \underbrace{\left(\frac{\kappa_{11}}{f_{11}^{\infty}}\right) f_{11} }_{f_{11}^{\rm poly}}+ \underbrace{\left(\frac{\kappa_{12}}{f_{12}^{\infty}}\right) f_{12}}_{f_{12}^{\rm poly}} + \underbrace{\left(\frac{\kappa_{22}}{f_{22}^{\infty}}\right)f_{22}}_{f_{22}^{\rm poly}}.
\label{eq:fpoly}
\end{eqnarray}
We continue to use Eq.~\ref{eq:div_simple} for the relative viscosity.

\begin{figure}[h]
	\centering
	\includegraphics[width=1.03\columnwidth]{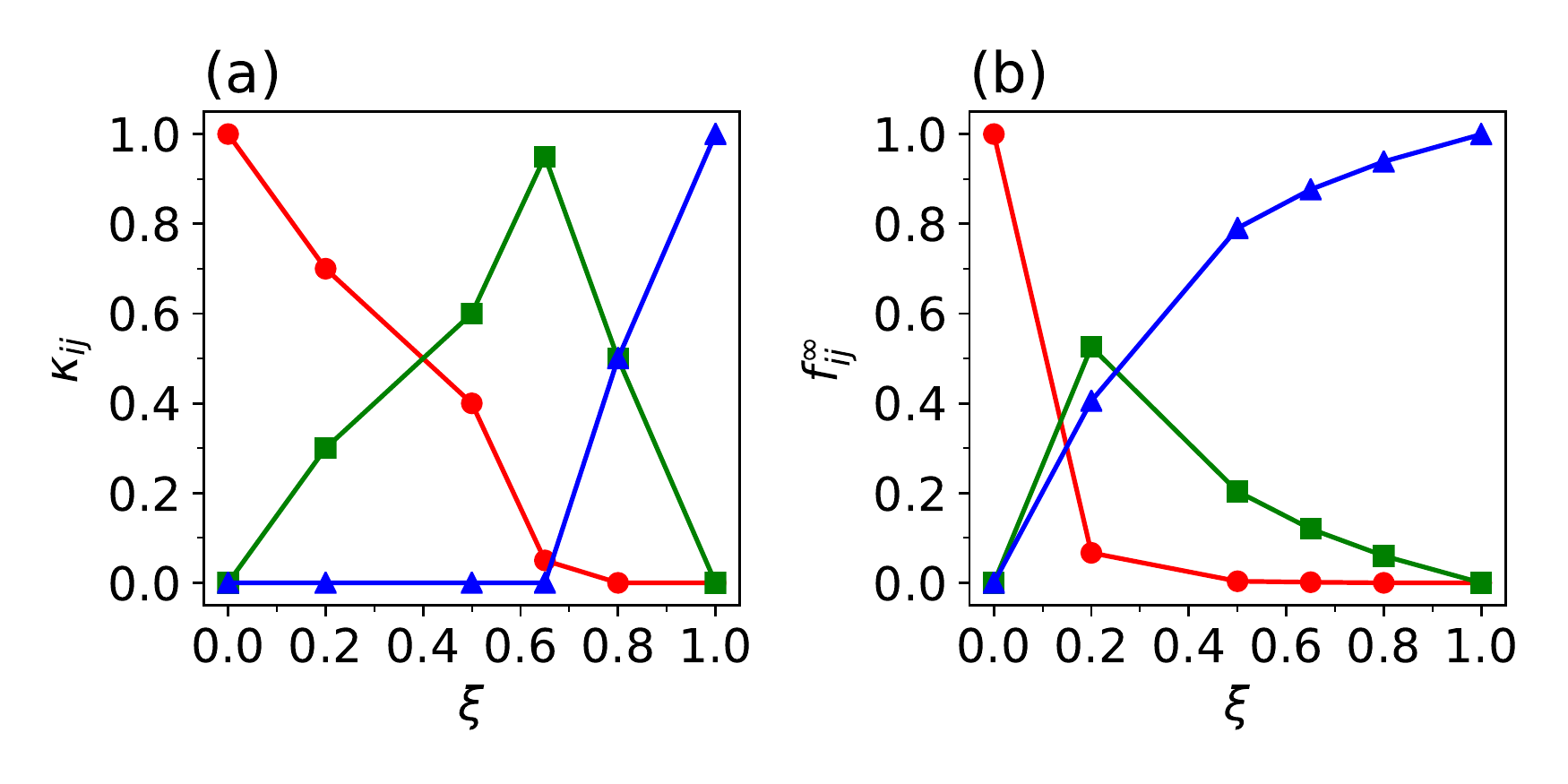}
\caption{(a) Coefficients, $\kappa_{11}$ (red), $\kappa_{12}$ (green) and $\kappa_{22}$ (blue), obtained from fitting bidisperse simulation flow curves using $f_{\rm poly}$, Eq.~\ref{eq:fpoly}, as a function of $\xi$.
(b) Fractions of frictional contacts in the large-stress limit, $f_{11}^{\infty}$ (11) (red), $f_{12}^{\infty}$ (green) and $f_{22}^{\infty}$ (blue), as a function of $\xi$.}
\label{fig:summary}
\end{figure}

The weighted fraction of frictional contacts of type $(ij)$, $f_{ij}^{\rm poly}(\sigma)$ needed to fit our data, shown in Fig.~\ref{fig:fij}(b) for $\xi=0.2$, has the same shape as $f_{ij}(\sigma)$, Fig.~\ref{fig:fij}(a), but scaled up by a factor of $\kappa_{ij}$, which sets the limiting value of $f_{ij}^{\rm poly}$ as $\sigma/\sigma_0 \longrightarrow \infty$.
[Note that the slight non-monotonicity of $f_{11}(\sigma)$ in Fig.~\ref{fig:fij}(a) (red) means that $f_{\rm poly}$ exceeds 1 using our normalisation scheme.]
By varying the free parameters $\kappa_{11}$ and $\kappa_{12}$ (with $\kappa_{22}=1-\kappa_{11}-\kappa_{12}$),  we can readily fit all of our bidisperse simulation flow curves; Fig.~\ref{fig:fij}(c) (solid line) shows the best-fit curve, obtained by eye, for $\xi=0.2$. In Fig.~\ref{fig:summary}(a), we plot the best-fit coefficients for $\xi=0.2$, along with the coefficients for other $\xi$ (for the full fits, see the ESI$^{\dag}$).

For comparison, we consider the special case in which $\kappa_{11}=f_{11}^{\infty}$, $\kappa_{12}=f_{12}^{\infty}$ and $\kappa_{22}=f_{22}^{\infty}$ , so that $f^{\rm poly}$ reduces to $f$, the unweighted fraction of frictional contacts, Fig.~\ref{fig:WC_fails}(a), and the original WC model is recovered, Fig.~\ref{fig:fij}(c) (dashed line). We plot the coefficients for this case, $f_{11}^{\infty}(\xi)$, $f_{12}^{\infty}(\xi)$ and $f_{22}^{\infty}(\xi)$, in Fig.~\ref{fig:summary}(b). 


For $\xi=0.2$, our set of fitted $\{\kappa_{ij}\}$, Fig.~\ref{fig:summary}(a), reaffirms quantitatively what we proposed qualitatively earlier. The largest contribution to changes in $f_{\rm poly}$ (and hence $\phi_{\rm J}$) is from large-large contacts, for which $\kappa_{11}=0.7$, while there is a negligible contribution from small-small contacts, $\kappa_{22}\approx 0$. In contrast, in the original WC model $\kappa_{11}=f_{11}^{\infty} \approx 0.1$ and $\kappa_{22}\approx 0.4$, Fig~\ref{fig:summary}(b). Increasing $\xi$ to $0.5$ sees the increasing importance of large-small contacts and decreasing importance of large-large contacts, while small-small contacts remain irrelevant. At $\xi=0.65$, $f_{\rm poly}$ is determined almost entirely by large-small contact formation. Only for $\xi=0.8$ do small particles have a measurable contribution, where they enter on equal footing with large-small contacts; large-large contacts are, by this point, irrelevant.

\section{Discussion}
\begin{figure}[t]
    \centering
    \includegraphics[width=0.7\columnwidth]{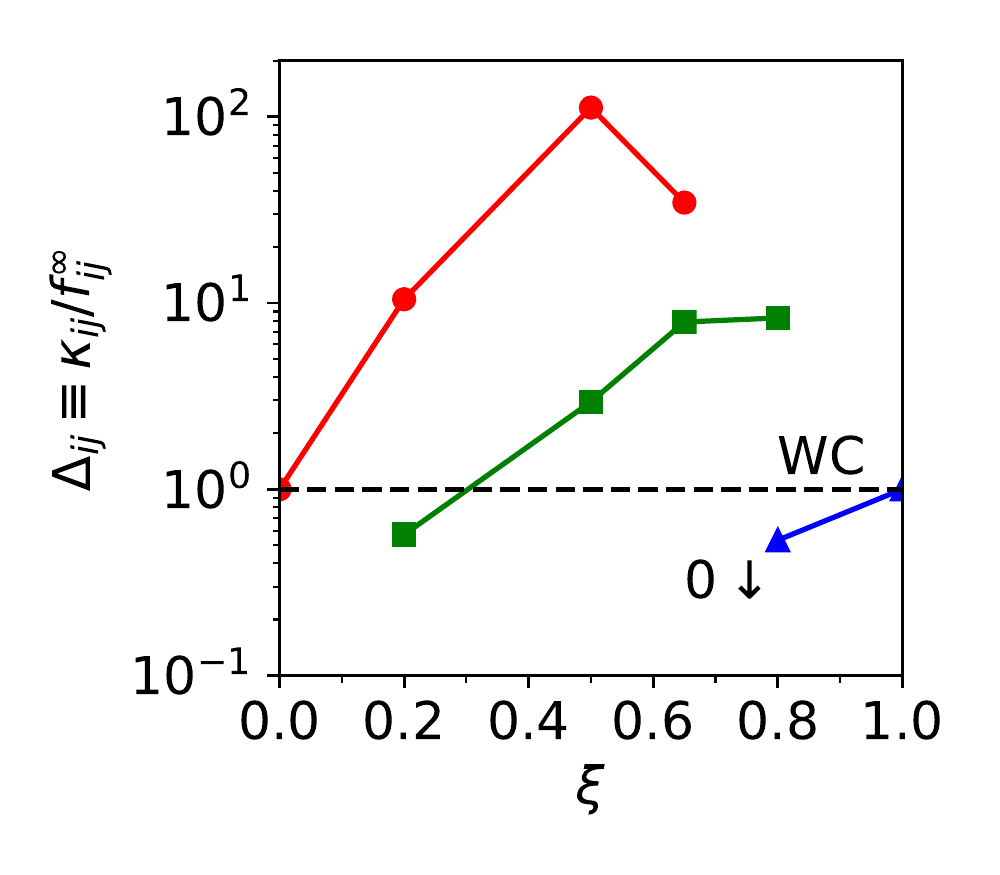}
    \caption{$\phi_{\rm J}$-contribution per contact $\Delta=\kappa_{ij}/f_{ij}^{\infty}$ for `11' (red), `12' (green) and `22' (blue) contacts, obtained by taking the ratio of the data in Fig.~\ref{fig:summary}(a) and (b). The horizontal dashed line is the WC prediction, $\Delta_{11}=\Delta_{12}=\Delta_{22}=1$.}
    \label{fig:relative}
\end{figure}

The WC coefficients, $f_{ij}^{\infty}$, in Fig.~\ref{fig:summary}(b) correspond to the relative numbers of each kind of contact (`11',`12' or `22') in the high-$\sigma/\sigma_0$ limit. Thus, the ratio $\Delta_{ij} \equiv \kappa_{ij}/f_{ij}^{\infty}$ measures the relative contribution to $\phi_{\rm J}^{\rm WC}$ (and hence to $\eta$) due to the formation of a single frictional contact of type $ij$. In the WC model, all contact types give rise to the same increment in $\phi_{\rm J}$ and $\Delta_{11}=\Delta_{12}=\Delta_{22}=1$. In Fig.~\ref{fig:relative}, we plot the $\Delta$s as a function of $\xi$ (colours) and overlay the WC prediction (horizontal dashed line). Strikingly, $\Delta_{11} \gg \Delta_{12} \gg \Delta_{22}$ for all our bidisperse mixtures. Thus,  at $\xi=0.2$, for example, a large-large contact contributes over an order of magnitude more than a large-small one; while, the effect of forming a small-small contact is negligible ($\Delta_{22} \approx 0$); i.e., small particles are largely redundant for stress tranmission. Only at $\xi=0.8$, where the fraction of small particles is largest and there are no large-large contacts ($\Delta_{11}$ is undefined here; so, we do not plot it), do small-small contacts have an appreciable contribution ($\Delta_{22}$ becomes non-zero). Even then, a single large-small contact contributes the same as $\sim\mathcal{O}(10)$~small-small contacts. 

Importantly, none of the bidisperse $\Delta$s follow the WC prediction (dashed line). Since different contact types do not contribute equally to changes in $\phi_{\rm J}^{\rm WC}$, one needs to know not only the total number of frictional contacts, but the sizes of the particles participating in all those contacts to predict $\eta$. Thus, $f$, which assumes that all frictional contacts contribute the same, is inherently unsuitable for modelling  bidisperse systems. 

The particle-size-dependence in Fig.~\ref{fig:relative} is reminiscent of sheared polydisperse dry granular packings, in which stress transmission is strongly spatially heterogeneous with contacts between larger particles carrying a higher load on average than those between smaller particles \cite{voivret2009multiscale}. In bidisperse dry granular systems, the detailed partition of stress is sensitive to both $\xi$ and the size ratio $\alpha$; e.g., as the size disparity grows ($\alpha$ decreases), the contribution of contacts involving small particles progressively decreases \cite{shire2016influence}. While this problem has been studied at length in dry systems under imposed particle pressure ($\implies$ varying $\phi$)\cite{gray2005theory,gray2006particle,weinhart2013discrete,cantor2018rheology}, it has received relatively little attention for fixed-$\phi$, granular suspensions, in which particles interact not only through contact-, but also hydrodynamic forces. Presumably, the trend with $\alpha$ is similar to the dry-grain case, so that, as $\alpha \rightarrow 1$, the disparity between different contact types vanishes, i.e., $\Delta_{11}\approx \Delta_{12} \approx \Delta_{22}=1$, and hence $f$ eventually becomes a reasonable approximation for $f_{\rm poly}$, which would explain the success of the WC model for weakly bidisperse mixtures\cite{singh2018constitutive}. Clearly, focussed work is needed to understand the details and origins of stress partitioning, and its relation to shear thickening, before a realistic model can be constructed. Minimally, the  the relative weights of different contact types should be allowed to vary, e.g., like our Eq.~\ref{eq:fpoly}.

Alternative existing models of frictional shear thickening could prove more successful in capturing bidisperse flowcurves. A recent idea is that thickening is not driven by the formation of frictional contacts \emph{per se}, but by the changes in anisotropy of stress tranmission that this induces\cite{thomas2018microscopic, otsuki2018shear}\footnote{This is reminiscent of the ``hydrocluster"-driven thickening of lubricated spheres observed in Stokesian dynamics simulations \cite{brady1985rheology}, which, however, is distinct from the contact-driven thickening we observe\cite{lin2015hydrodynamic}}. 
In particular, Thomas~\emph{et. al.} proposed an \emph{ab initio} model for two-dimensional systems based on the ratio of the shear stress and particle pressure, $\sigma_{xz}/P$, which they relate to the anistropy of contact forces. Interestingly, for simulations of dry grains with a uniform particle size distribution, $\sigma_{xz}/P$ is found to be \emph{independent} of polydispersity (controlled via the span of the distribution) \cite{voivret2009multiscale,cantor2018rheology}, which suggests that their approach may account for polydispersity effects naturally in a way that the WC model, which is agnostic to the spatial distribution of contact forces, does not. This merits a thorough study of the role of stress anisotropy in bidisperse systems, and the extension of Thomas~\emph{et. al.}'s theory to 3-d and polydisperse systems.

Perhaps as important as the contact-type-dependent contributions to $\phi_{\rm J}$ is our observation that the original WC model fits our monodisperse simulation data. This result is non-trivial: it implies that the microphysics of shear thickening can be captured by a single scalar parameter ($f$) that is agnostic to the spatial distribution of contacts. Remarkably, there is evidence in the literature that this scenario may be true, at least in the quasi-monodisperse limit. By simulating a weakly bidisperse mixture of particles interacting via the critical-load model and the same mixture containing particles with different but load-independent $\mu$, Dong and Trulsson\cite{dong2017analog} showed that $\eta$ is uniquely defined by $f$, even though the microstructure for both setups is very different. For strongly bidisperse suspensions, the roles of microstructure and stress paritioning in shear thickening remain to be disentangled. If they bear similarity to polydisperse dry grains, then the two should be strongly correlated. Specifically, we would expect big particles, which carry the largest loads, to align with the compressive axis; whereas small particles, which carry a negligible load, would form an almost isotropic background of ``spectator" particles\cite{voivret2009multiscale,radjai1998bimodal}. Studying the spatial distribution of contacts systematically in these systems, e.g., in the vein of Dong and Trulsson\cite{dong2017analog}, would shed light on this issue and help to establish whether the independence to microstructure in the monodisperse case has deeper physical meaning, or if it is entirely fortuitous.

Before concluding, we comment on the utility of experimental data in testing the WC model. Although our experiments qualitatively support the notion that the WC model fails for bidisperse suspensions, the inability of the model to decribe the all-small $\eta(\sigma)$ based on simulation $f(\sigma)$ highlights an important obstacle to rigorous testing: the $F$-dependent tribology of interparticle contacts is \emph{a priori} unknown. Our work with PHSA-stabilised PMMA dispersions suggests that, even for particles with ostensibly well-controlled surface properties, $\mu(F)$ may vary considerably from batch to batch. For example, in the ESI$^\dag$ we show flow curves for quasi-monodisperse PMMA spheres showing ``two-stage'' shear-thickening, with two distinct onset stresses. Such behaviour clearly cannot arise from the single-stress-scale CLM used here. Thus, microtribology experiments\cite{comtet2017pairwise,galvez2017dramatic,james2019tuning} must play a central role in future model testing. Indeed, the scarce tribology measurements that already exist for sterically-stabilised particles  indicate a rich behaviour, particularly at large normal loads\cite{chatte2018shear}. Finally, we note that even in experimental systems where $\mu(F)$ is described by the CLM, ``fitting'' the WC model to experimental data with a presumed form for $f(\sigma)$ will result in a $f(\sigma)$ that is \emph{not} the fraction of frictional contacts except in the monodisperse limit\cite{guy2015towards,royer2016rheological}.

\section*{Summary and conclusions}
We have simulated and experimentally measured the rheology of a bidisperse suspension of repulsive spheres to test the validity of the WC model of shear thickening. By using the fraction of frictional contacts $f$ extracted directly from simulations, we showed that the WC model works in the special case of monodisperse particles, but is incomplete when applied to bidisperse mixtures. While our study focusses on continuous shear thickening, we expect all the same conclusions to apply at higher volume fractions, where discontinuous shear thickening is observed.

In practical terms, our results suggest caution when using the WC model as anything other than an empirical fitting tool. Specifically, little, if any, meaning can be ascribed to $f$ extracted from fits to flow curves.
On a fundamental level, our work highlights the need for a focussed effort to understand the link between $\sigma$-dependent frictional contact formation and dissipation. Existing studies of shear thickening consider either bulk rheology\cite{guy2015towards,royer2016rheological} or \emph{ex-situ}, two-particle properties\cite{comtet2017pairwise}, with little or no concerted effort to bridge the two regimes. We believe that any serious effort to make this link should consider polydispersity from the outset in its own right, rather than merely as a means of mitigating crystallisation. Indeed, our work hints that the monodisperse limit is a singular one, and so cannot be used as a guide to developing models for the flow of polydisperse systems. 

Data from this article are available at https://doi.org/10.7488/ds/2644.

\section*{Conflicts of interest}
There are no conflicts to declare.

\section*{Acknowledgements}
BMG and MH were funded by EPSRC EP/J007404/1. CJN was funded by EPSRC EP/N025318/1 and the Maudslay-Butler Research Fellowship at Pembroke College, Cambridge. LJS was funded by EPSRC SOFI CDT (EP/L015536/1). JS was funded by EPSRC EP/N025318/1 and The Royal Academy of Engineering/The Leverhulme Trust Senior Research Fellowship LTSRF1617/13/2. WCKP was funded by EPSRC EP/J007404/1 and EP/N025318/1. We thank Andrew Schofield for synthesising the particles, and John Royer, Dan Hodgson and an anonymous referee for helpful discussions. The simulation makes use of the LF-DEM code published in Mari \emph{et. al.}\cite{mari2014shear} and available at \url{https://bitbucket.org/rmari/lf_dem} as well as LAMMPS \cite{Plimpton1995}. 


\balance


\bibliography{arXiv_SoftMatter_Binary} 
\bibliographystyle{rsc} 

\end{document}